\documentclass[12pt,preprint,notoc,nohyper]{MCsComment} % 10pt is ignored!

\usepackage{epsfig,multicol,bbm}

\newcommand\fverb{\setbox\pippobox=\hbox\bgroup\verb}
\newcommand\fverbdo{\egroup\medskip\noindent%
            \fbox{\unhbox\pippobox}\ }
\newcommand\fverbit{\egroup\item[\fbox{\unhbox\pippobox}]}
\newbox\pippobox
\title{Comments on matter collineations of plane symmetric, cylindrically symmetric
and spherically symmetric spacetimes}

\author{Asghar Qadir$^{a}$ and K. Saifullah$^b$ \\

$^a$Centre for Advanced Mathematics and Physics, National
University of Sciences and Technology, Rawalpindi, Pakistan \\
$^b$Department of Mathematics, Quaid-i-Azam University, Islamabad,
Pakistan  (Electronic address: aqadirmath@yahoo.com,
saifullah@qau.edu.pk) \\}

\preprint{}  % OR: \preprint{Aaaa/Mm/Yy\\Aaa-aa/Nnnnnn}
\abstract{Comments are made on some recently published papers on
matter collineations of plane symmetric, cylindrically symmetric and
spherically symmetric spacetimes.}

\dedicated{}

\begin{document}

Recently matter collineations (MCs) of plane symmetric static \cite
{sharifpln} and cylindrically symmetric static spacetimes
\cite{sharifcyl} have been presented. Earlier, the same author also
classified spherically symmetric static spacetimes according to
their MCs \cite{sharifsph}. For an energy-momentum tensor,
$\mathbf{T}$, we call $\mathbf{\xi }$ an MC if
\begin{equation}
{{\Large \pounds }_{\mathbf{\xi }}}\mathbf{T}=0\;. \label{mc}
\end{equation}
In component form, Eq. (\ref{mc}) becomes the MC equation
\[
T_{ab,c}\xi ^{c}+T_{ac}\xi _{,b}^{c}+T_{bc}\xi _{,a}^{c}=0\;.
\]
In these equations, if $\mathbf{T}$ is replaced by the Ricci tensor, $%
\mathbf{R}$, then the vector $\mathbf{\xi }$ is called a Ricci
collineation (RC). Noting the apparently similar form of the
equations and the role of the matter and Ricci tensors in the
Einstein field equations (EFEs)
\begin{equation}
R_{ab}-\frac{1}{2}Rg_{ab}=\kappa T_{ab} ,  \label{1.111}
\end{equation}
the author merely replaced the Ricci tensor by the matter tensor in
Refs. \cite{FQZ}-\cite{mziad} for the classification according to
RCs of plane symmetric, cylindrically symmetric and spherically
symmetric static spacetimes, respectively. All that remains to be
done is to check that the RCs satisfy the MC equations. Even if he
has done so, errors persist in his papers. For example, because of
the error in calculations the Lie algebra for a case, as given in
Eq. (B47) of Ref. \cite{FQZ}, does not close. This error has been
carried over in Eqs. (43) of his paper \cite{sharifpln}. (A
typographical error there is carried over as well.) Now, as plane
symmetry can locally be considered as a special case of cylindrical
symmetry, this particular case appears in Ref. \cite{QSZ} also,
where it has been corrected. This correction has been carried over
into Eqs. (27) of his paper \cite{sharifcyl} as well, but the author
has not cited these papers!

It is worth mentioning a serious misconception, in the three subject
papers \cite{sharifpln, sharifcyl, sharifsph}, that was not imported
from the Ricci collineation papers. The author says that there are
``three, four, five, six, seven or ten MCs out of which three are
isometries and the rest are proper'' \cite{sharifpln, sharifcyl}. He
has assumed that the isometry group is minimal. This is simply
incorrect as there are numerous cases of non-minimal isometry
groups. It is possible that all the four, five, six, seven or ten
MCs may be isometries and there may be no proper MCs. A similar
problem arises for the spherically symmetric case \cite{sharifsph}.

We also mention here that Ref. \cite{saifpln} on the same subject as
Ref. \cite{sharifpln} does not only classify the plane symmetric
spacetimes (with the correct Lie algebras provided) but it discusses
the issue of the relationship between the RCs and MCs and provides a
number of explicit examples for that purpose also. We will not go
into further detail on this because this is a subject of a separate
full-length study \cite{KQSZ} in itself.


\begin{thebibliography}{999}


\bibitem{sharifpln}  Sharif, M., \textit{J. Math. Phys.} \textbf{45 }(2004)
1518.

\bibitem{sharifcyl}  Sharif, M., \textit{J. Math. Phys.} \textbf{45 }(2004)
1532.

\bibitem{sharifsph}  Sharif, M., \textit{J. Math. Phys.} \textbf{44 }(2003)
5141.

\bibitem{FQZ}  Farid, T. B., Qadir, A., and Ziad, M., \textit{J. Math. Phys.}
\textbf{36 }(1995) 5812.

\bibitem{QSZ}  Qadir, A., Saifullah, K., and Ziad, M., \textit{Gen. Rel.
Grav. } \textbf{35 }(2003) 1927.

\bibitem{mziad}  Ziad, M., \textit{Gen. Rel. Grav. }\textbf{35 }(2003)
915.

\bibitem{saifpln}  Saifullah, K., arXiv: gr-qc/0406095.

\bibitem{KQSZ}  Khan, H., Qadir, A., Saifullah, K., and Ziad, M., \textit{In
preparation}.

\end{thebibliography}
\end{document}